# Controversy in Evolutionary Theory:
# A multilevel view of the issues.

**George Ellis**[1]

**Mathematics Department,**
**University of Cape Town**

**Abstract:** *A conflict exists between field biologists and physiologists ("functional biologists" or "evolutionary ecologists") on the one hand and those working in molecular evolution ("evolutionary biologists" or "population geneticists") on the other concerns the relative importance of natural selection and genetic drift. This paper is concerned with this issue in the case of vertebrates such as birds, fishes, mammals, and specifically humans, and views the issue in that context from a multilevel perspective. It proposes that the resolution is that adaptive selection outcomes occurring at the organism level chain down to determine outcomes at the genome level. The multiple realizability of higher level processes at lower levels then causes the adaptive nature of such processes at the organism level to be largely hidden at the genomic level. The discussion is further related to the "negative view" of selection, the Evo-Devo and Extended Evolutionary Synthesis views, and the levels of selection debate, where processes at the population level can also chain down in a similar way.*

## 1: Introduction

There are currently a number of unresolved major controversies in evolutionary theory. Apart from the ongoing conflict about sociobiology, evolutionary psychology, and all that, which I will not discuss here,[2] these include,

1. **Drift/selection**: A conflict between field biologists and physiologists ("functional biologists" or "evolutionary ecologists") on the one hand, and those working in molecular evolution ("evolutionary biologists" or "population geneticists") on the other (Mayr 1961, Birch 2016), with the latter (e.g. Lynch 2007) vehemently claiming that drift is almost always more important than selection (at least at the molecular level).
2. **The "negative view"** suggesting selection cannot explain the origin of traits (Stegmann 2010, Birch 2012);
3. **Evo-devo/the Extended Evolutionary Synthesis (EES)** versus **Standard Evolutionary Theory (SET),** as discussed by Laland, Uller, *et al* (2014) in debate with Wray *et al* (2014);
4. **The levels of selection issue:** can selection take place at the population level? (Lewontin 1970, Okasha 2006). This is the kin selection/inclusive fitness vs group selection/multilevel selection debate (Birch and Okasha 2015, Wade *et al* 2010, Kramer and Meunier 2016).

My main concern in this paper is the first conflict, which is evident particularly in polemical blogs such as those by P Z Myers (e.g. Myers 2014) and L Moran (e.g. Moran 2016), in opposition to views held by virtually all functional biologists. The two views on the relative importance of selection and drift result in a strange contrast. Books such as the major biology text by Campbell and Reece (2005) take it for granted that natural selection leads to adaptation to the environment, as does the animal physiology text by Randall *et al* (2007), the psychology text by Gray (2011), and the ground-breaking book *Arrival of the Fittest* by Wagner (2017). A typical quote is

> *``The physiology of an animal is usually very well adapted to the environment that the animal occupies, thereby contributing to its survival. Evolution by natural selection is the accepted explanation for this condition, called adaptation. .. A physiological process is*

[1] Email: george.ellis@uct.ac.za.
[2] For a detailed proposal in this regard, see Ellis and Solms (2017).

*adaptive if it is present at high frequency in the population because it results in a higher probability of survival and reproduction than alternative processes"* (Randall *et al* 2002:7)

Similar statements will be found in each of the above books, and in texts on evolutionary theory such as by Mayr (2002) and Kampourakis (2014). In an interview in 1999,[3] Mayr stated

"*Darwin showed very clearly that you don't need Aristotle's teleology because natural selection applied to bio-populations of unique phenomena can explain all the puzzling phenomena for which previously the mysterious process of teleology had been invoked".*

By Contrast, Lynch writes in the preface to his major book *The Origins of Genome Architecture* (Lynch 2007: xiii-xiv, quoted in Moran 2016):

*Contrary to popular belief, evolution is not driven by natural selection alone. Many aspects of evolutionary change are indeed facilitated by natural selection, but all populations are influenced by non-adaptive forces of mutation, recombination, and random genetic drift. These additional forces are not simple embellishments around a primary axis of selection, but are quite the opposite—they dictate what natural selection can and cannot do … A central point to be explained in this book is that most aspects of evolution at the genome level cannot be fully explained in adaptive terms, and moreover, that* ***many features could not have emerged without a near-complete disengagement of the power of natural selection****. This contention is supported by a wide array of comparative data, as well as by well-established principles of population genetics"* [my emphasis],

and see also Lynch (2007a). Similarly Myers (2014) states

"*First thing you have to know: the revolution is over. Neutral and nearly neutral theory won. The neutral theory states that* ***most of the variation found in evolutionary lineages is a product of random genetic drift.*** *Nearly neutral theory is an expansion of the idea that basically says that even slightly advantageous or deleterious mutations will escape selection — they'll be overwhelmed by effects dependent on population size. This does not in any way imply that selection is unimportant, but only that* most *molecular differences will not be a product of adaptive, selective changes*" [my emphasis].

For skeptical responses to Lynch's book, see Pigliucci (2011) and Charlesworth (2008). The question is, how there can be a compatibility between these two strongly held views? If most of the variation found in evolutionary lineages is a product of random genetic drift, how does apparent design arise? It surely can't be an accidental by-product of random events – that was the whole point of Darwin's momentous discovery (Darwin 1872) of a mechanism to explain apparent design that is so apparent in all of nature (Mayr 1961,2002). On the face of it, Lynch, Myers, and Moran seem to be saying the ID people are right: evolution cannot adapt life to its environment, because random effects dominate.

This paper proposes that in order to reconcile these apparently opposing views, one must look at the issue explicitly in the light of the levels of emergence involved; that is, this issue cannot be resolved separately from the multilevel debate. Figure 1 shows the broadest relevant levels: the level of genes **L1**, the level or organisms **L2**, the level of populations **L3,** and the environment level **L4**. Basically in the end there are three types of multilevel relations to be considered: **ML1**, that between the organism and gene; **ML2**, that between the population and the organism, and **ML3**: that between the environment and the population. How does selection relate to these levels? Note that as stated in the abstract, the specific concern of this paper is cases such as finches, fishes, giraffes, vertebrates in general, and particularly humans. All the studies of evolutionary processes in wheat, *C elegans*,

[3] E Mayr (1999): https://www.edge.org/conversation/ernst_mayr-what-evolution-is

*Drosophila*, and so on provide important information on the processes at work, but do not deal with the full complexity of what happens in the cases of vertebrates in general, and in particular animals with complex social structures.

The proposal here, as explicated below, is that selection takes place at either the individual level **L2** (Section 3) or population level **L3** (Section 5), and this outcome then chains down to the level **L1** of gene regulatory networks, genes, and proteins. Indeed that must be how individuals and populations get adapted to the environment; as otherwise such adaptation is unexplained. Because both RNA and the proteins involved in gene regulatory networks as well as all other proteins needed in the cell are encoded by DNA, in effect an image of the environment is imbodied in the genome (Stone 2015:188).

---

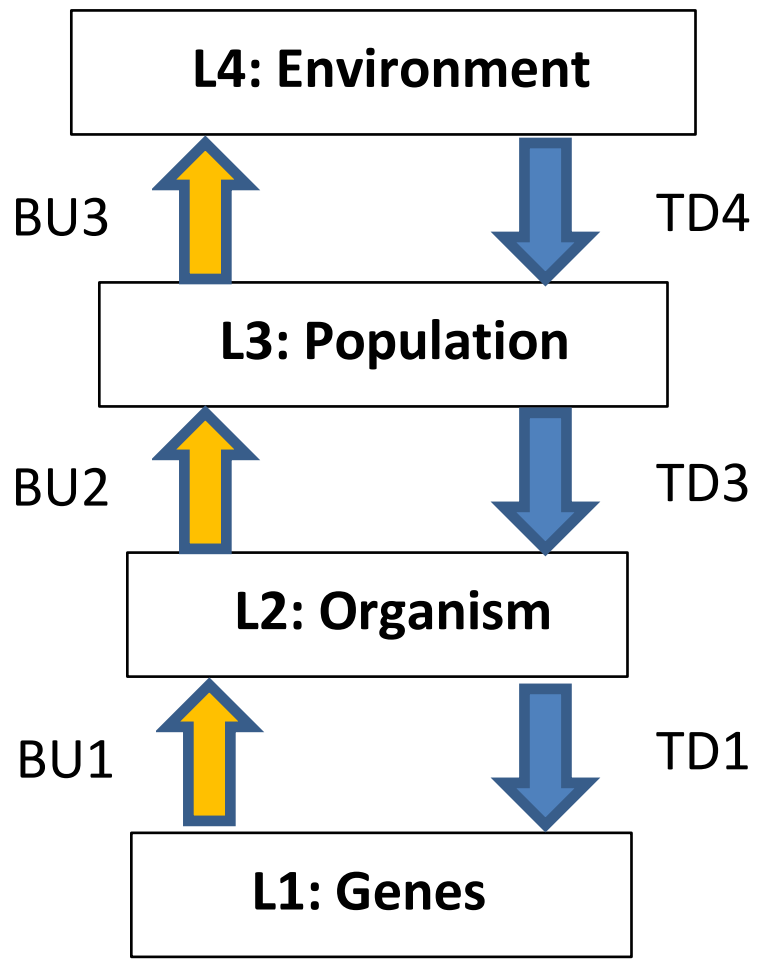

**Figure 1: The broad levels involved in evolutionary selection.** *Bottom up emergence BU1, BU2, BU3 and top-down effects TD4, TD3 and TD1 all occur, as discussed in this note.*[4] *More detailed diagrams of relevant effects are given below; the total set of interactions is summarised in Figure 7.*

---

Between **L3** and **E** one has the issue of niche construction (Laland *et al* 2000). Between **L2** and **L3** one has the kin and multilevel selection debate proper (Wade at el 2010, Birch and Okasha 2015, Kramer and Meunier 2016). Between **L1** and **L2** one has the contrast of the view from population genetics involving genetic drift at level **L1** (Lynch 2007, 2007a) versus the view from field biology and physiology at level **L2** (Rhoades and Pflanzer 1989, Randall *et al* 2002, Campbell and Reece 2005). It will be the contention of this paper, in agreement with Schaffner (1998), Martinez and Moya (2011) and Noble (2017), first that it is crucial to look at these opposing views in a multi-level context (Figures 1-7), and second that as well as bottom-up emergence **BU1, BU2** and **BU3** between these levels, there are also top-down effects **TD4**, **TD3** and **TD1** occurring between them[4] that crucially shape outcomes (Noble 2008, 2009, 2012, 2016, Ellis *et al* 2012, Ellis 2016). In a nutshell, selection effects are clear at levels **L2** or **L3**, because that is where they usually occur, but are somewhat hidden at level **L1**, because that is almost always not the level at which they occur. Such effects can indeed explain apparent design at level **L2**: the Darwinian project is not undermined by population genetics.

The stage for the discussion is set in Section 2 by considering the relation between structure, function, and traits. The drift/selection issue is discussed in Section 3 in terms of the relation between **L1** and **L2.** This section also looks at the "negative view" of selection. Section 4 gives a more complex view of the relations involved as proposed by Evo-Devo and "Extended Evolutionary Synthesis" (EES) views of these processes, while Section 5 considers the multilevel selection issues as normally understood: that is, the relation of **L2** and **L3** to each other and to **L4**. The overall pattern of causation emerging, and its relation to the EES, is discussed in Section 6.

---

[4] **TD2** is the direct top down effect of the environment on organisms, see Figures 2 and 6.

## 2: Structure, Function, and Traits

As a preliminary, one must ask what kind of things might be selected for: are they structures, functions, traits, or what? In studies of physiology (Rhoades and Pflanzer 1989, Randall *et al* 2002) and integrated views of biology (Campbell and Reece 2005), these are intimately intertwined

One cannot sensibly talk about physiology of living systems without talking about function or purpose: the heart exists in order to circulate blood (Randall *et al* 2002:476-510), pacemaking cells exist in order to determine the rhythm of the heart (Noble 2002), blood exists in order to transport oxygen, a key function of mitochondria in eukaryotes is to provide energy to the cell by converting ingested sugars into ATP (Randall *et al* 2002:74), and so on (Rhoades and Pflanzer 1989). Thus in *Animal Physiology*, Randall *et al* (2002:3,6) write

> *Animal physiology focuses on the functions of tissues, organs, and organ systems in multicellular animals... Function flows from structure... A strong relationship between structure and function occurs at all levels of biological organisation."*

Functional talk in biology is usually taken to be vindicated by Darwinism, as discussed *inter alia* by Ernst Mayr (2002,2004) and Kampourakis (2014). According to Kampourakis,

> *``Function is the role of a component in the organization of a system. The functions of the parts and activities of organisms in enabling their continued existence are ``biological functions" or ``biological roles" (the function of the wing of an eagle is to enable flight, but the function of the wing of a penguin is enabling swimming)"* (Kampourakis 2014:221)

A more formal related definition is given by Farnsworth *et al* (2017). According to Godfrey-Smith (1994), supported by Birch (2017:22),

> *``Biological functions are dispositions or effects a trait has which explain the recent maintenance of the trait under natural selection".*

According to Randall *et al*,

> *``The physiology of an animal is usually very well adapted to the environment that the animal occupies, thereby contributing to its survival. Evolution by natural selection is the accepted explanation for this condition, called adaptation. .. A physiological process is adaptive if it is present at high frequency in the population because it results in a higher probability of survival and reproduction than alternative processes"* (Randall *et al* 2002:7).

In this note, following Godfrey-Smith(1994), Mayr (2002, 2004), Kampourakis (2014), Noble (2017), and Birch (2017), the view is as follows:

> The **function** $\boldsymbol{\alpha}$ of a trait **A**, or of a physiological system **S** that enables **A**, exists in individuals in a population because having property $\boldsymbol{\alpha}$ tends to increase differential reproduction and survival rates for a population of these individuals, and so natural selection (Darwin 1872) is a mechanism for developing property $\boldsymbol{\alpha}$ over recent evolutionary timescales, resulting in genotypes **X** that will lead to existence of property $\boldsymbol{\alpha}$ in individuals through contextual developmental processes. Selection takes place simultaneously for the function, trait, and underlying physiological systems that enable them.

Thus proper function of a trait is the effect for which it was selected by natural selection (Neander 1991). Note that this is a definition at level **L2.** Farnsworth *et al* (2017).show how this then leads to a definition of function at a lower level:

> **A biological function** *is a process enacted by a biological system A at emergent level n which influences one or more processes of a system B at level n + 1, of which A is a component part*

This defines function at level **L1** from that at level **L2** if we set n = 1.

**3: Drift and level of selection**

The drift controversy about the nature of biological evolution is due to the way many evolutionary theorists work essentially at the genetic level, where it is claimed that selection plays a minor role in what is going on. But for the multicellular organisms which are the concern of this paper, selection has no handle whereby it can directly effect what happens at the gene level (Mayr 2002, Martinez and Moya 2011). Rather, selection takes place at the organism level,[5] due to whether they survive or die before they have had time to reproduce, that selection then chaining down to the genotype level (organisms that survive pass on their genes to the next generation).

This is indicated in Figure 2, where the main chain of causation is indicated by the yellow arrows. The overall effect (Natural Selection, **NS)** is necessarily a multi-level process.

The currency of selection is expected lifetime reproduction success (total number of offspring surviving to reproductive stage). This is the reproductive potential: expected reproduction rate / expected mortality rate, as used in life-history optimisation.

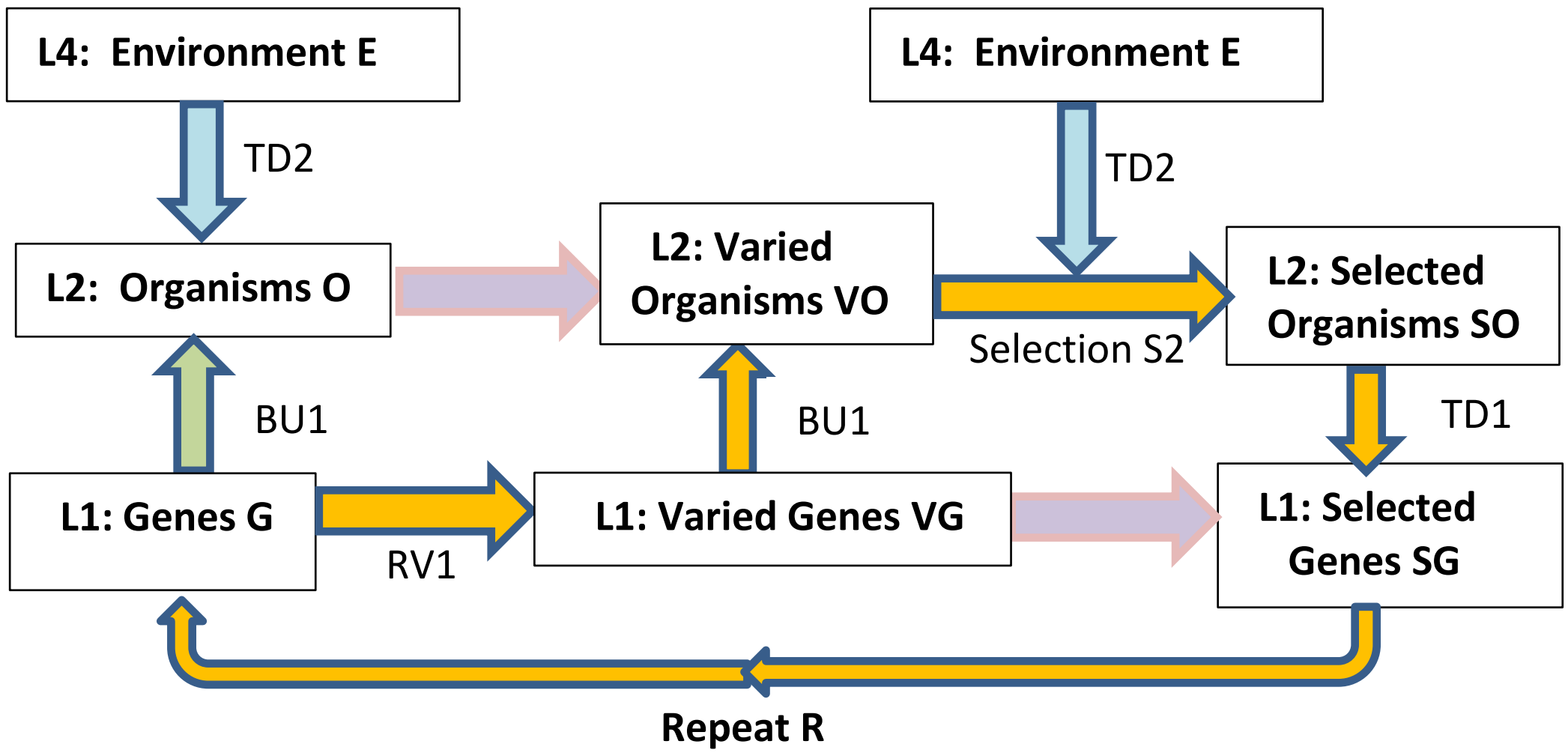

**Figure 2: The basic causal chain whereby selection leads to a new set of genes.** *The yellow arrows with dark edges show the causal mechanisms operating on the initial set of genes* **G**, *the light purple ones show the resultant effective outcomes. The blue arrows with dark edges show the top-down effect of environmental constraints on selection. The selected genes* **SG** *form the starting point for the next round of selection.*

[5] Or possibly even at the population level (see Section 5)

### 3.1 The basic multilevel process

The genotype **G** at level **L1** can specify either proteins or functional RNA, and leads through contextually dependent developmental processes **BU1** to individual organisms **O**, which have to be of a kind permitted by the environment (e.g. they can only be oxygen-breathers if the atmosphere contains oxygen; this will be satisfied because of the previous rounds of selection)**.** This is the top-down influence **TD2**. Replication with Variation **RV1: G➔ VG** takes place at the genotype level in various ways; this is where much randomness enters. The varied set of genes **VG** lead through the same developmental processes **BU1** to a varied set of organisms **VO** at the phenotype level, with random variations. Selection **S2**: **VO ➔ SO** then takes place at that level; variations may found to be neutral, deleterious, or advantageous in the environmental context **E**. A subset of organisms for which this variation is at least not too deleterious will survive, resulting in a selected set of individuals **SO** generally better suited to the environment **E**, or at least not too much worse adapted. This necessarily results in the survival only of a subset **SG** of the varied genes **VG** (because they are the genes embodied in the surviving organisms **SO**). This is the top-down process **TD1** that determines **SG,** which are the basis for the start of the next round (Campbell 1974, Murphy and Brown 2005). Varying the environment **E** leads to different outcomes **SO** and **SG** of the whole process; this is the top-down effect **TD2**. Finally the repeat operation **R** uses the selected set of genes **SG** for the start of the next round of variation and selection.

The overall process **NS1** ("Natural Selection") for one round at the genotype level is (see Figure 2),

**NS1: G➔ SG = {RV1 ➔ BU1 ➔ S2 ➔ TD1**}. (1)

The overall outcome at the organism level **L2** is a map **O ➔ SO**, characterised by selection leading by and large to relatively improved adaptation to the environment **E**. The **L2** selection process **S2** (a projection operator, see Ellis and Kopel 2017) is the process by which individuals are culled leading to adaptive outcomes at level **L2**:

**S2: VO➔ SO** **(2)**

leading to natural selection **NS2** as seen at the organism level:

**NS2: O ➔ SO** **(3)**

If one looks only at the genotype level **L1**, one sees only the map

**VG: G➔ SG** **(4)**

as the effective outcome, not directly characterised by selection but rather being an indirect result of selection (2) at the level **L2**, as indicated by (1). The process is no longer obviously adaptive because of the multiple realisability of phenotypes in terms of their underlying genotypes (Wagner 2017): there is a very large equivalence class of genes **SG** that can lead to the class of individuals **SI** that survive the selection process **L2** and so pass their genes on to their progeny at the start of the next round. In general the selection process at level **L2** does not lead to simple outcomes at level **L1**.

**Issues arise** at each step in this causal chain (Neander 1988, Mayr 2002, Lynch 2007a, Birch 2012, Noble 2013, Birch 2013, Scholl and Pigliucci 2014, Laland, Uller, *et al* 2014, Noble *et al* 2014, Birch 2016, 2018), as follows (see Figure 2):

- **Replication with Variation RV1** involves very complex processes (Kirkpatrick *et al* 2002, Lynch and Conery 2003, Lynch 2007, Noble *et al* 2014) of both genetic and epigenetic nature. There is a delicate balance between not enough variation and too much variation

- **at the genotype level**, how does this variation take place? By mutation, recombination, gene duplication, and mobile genetic elements (lateral gene transfer including symbiogenesis) (Noble *et al* 2014)
- **as regards population genetics**, the outcome is affected by genetic drift , which is the change in the frequency of an allele in a population due to random sampling of organisms (Lynch 2007**)**, see \S3.3
- **What is the role of chance?** Is the process to some degree directional? **(**Laland Uller *et al* 2014, Noble and Noble 2017). This is a very controversial area, see \S4.3.

- **Developmental Processes BU1** (West-Eberhard 1989, 2003, Oyama *et al* 2001,Pigliucci 2010, Noble *et al* 2014, Pigliucci 2017) are immensely complex processes They involve
  - **Self-assembly processes (**Kauffman 1992, 1993) represented by Fitness Landscapes in Sequence Space, the NK Model of Rugged Fitness Landscapes, and the NK Model of Random Epistatic Interaction
  - They crucially involve **Developmental Systems** as actors (Oyama *et al* 2001, Carroll 2005, Gjuvsland *et al* 2013, Noble 2008a, Mattick 2012, Noble *et al* 2014). This key feature is discussed in Section 4.1, leading to the more complex view in Figure 4.
  - **Developmental bias:** these processes generate certain forms more readily than others (Laland Uller *et al* 2014, Laland et al 2015), leading to convergent evolution (Conway Morris 2003, McGhee 2011). This is partly because only those organisms are possible, and partly because it is easier to generate some forms than others.
  - **Developmental plasticity** (West-Eberhard 1989, 2993) also occurs: an environmental sensitivity through epigenetic effects (Gilbert and Epel 2009, Noble 2015) mediated by gene regulatory networks (Carroll 2008), see \S4.1. This is a crucial effect.
- **Physiology L2:** what can arise is limited by what is physiologically possible (Noble 2013).
  - **Convergent evolution (**Conway Morris 2003, McGhee 2011) is a consequence: evolution seeks out the limited sets of possible ways to solve functional problems.
  - These are related to **contextual constraints** set by the current environmental context **E**, see e.g. Ballance (2016), Wagner (2017), Godfrey-Smith (2017) for specific examples. This is thus the constraining nature of the environment **TD2**.
- **Selection S** deletes unsuccessful individuals in a population, with result they die without reproduction. The outcome is the restriction **V0➔ SO** of varied organisms to selected ones that successfully reproduce, which thereby alters both the composition of the population (the bottom up effect **BU2** in Figure 5) and of the gene pool (the top down effect **TD1**). This is the way that a more ordered population **SO** is generated at level **L2** from the varied population **VO,** in general more suited to the environment, or at least no less fit**.**
  - Selection at the lower levels by this means is in fact for **developmental systems** and the molecules they require to function, not for genes *per se* (Oyama *et al* 2001, Pigliucci 2010, Wagner 2017). This is a crucial feature, see \S4.1
  - **Contextual effects TD2** govern selection: the environment creates niches that define what is possible and so govern selection outcomes. Different environment result in different organisms being possible (Ballance 2016, Godfrey-Smith 2017, Wagner 2017)
  - The most accurate way to describe these effects is via a "**norm of reaction**" approach (Lewontin 1995) which is " list or graph of the correspondence between different possible environments and the phenotypes which would result" (Schaffner 1998)
- **Realisation at cell and molecular level TD1**: selection chains down from the selected organisms to the level of developmental systems and the needed DNA**,** RNA and proteins for them to function (Petsko and Ringe 2009, Wagner 2017, Ellis and Kopel 2017).
  - This is the top-down realisation **TD1** of selection **S2** occurring at level **L2,** to level **L1.**
  - A key feature here is the multiple realisability of phenotypes in terms of genotypes (Wagner 2011, 2017). This is discussed in \S3.4

Apart from epigenetic effects, dealt with in Section 4, three crucial issues arise in these processes: population effects leading to drift (\S3.2), multiple realisability (\S3.3), and existence of many more levels (\S3.4). The negative view is discussed in \S3.5, and the overall outcome in \S3.6.

### 3.2 Population effects: drift

The diagrams do not attempt to represent the relation of the individuals to the populations they belong to. There is a huge literature on this relationship (see Lynch 2007, Birch 2013, Birch and Okasha 2015, Kramer and Meunier 2016, Okasha 2016, Walsh et al 2017, Birch 2018, and references therein), which is key because selection acts on populations through its effects on individuals: it can't act on a population *per se* (Martinez and Moya 2011). It acts at level **L2** through survival of individuals and then chains up to **L3** and down to **L1**, see Figure 1. Thus one must deal with explanation of traits of individuals as well as the distribution of traits in a population **(**Neander 1988). This relationship is the source of the level of selection issue, see Section 5, and of genetic drift, as discussed by Lynch (2007).

**Meaning of Drift**: Pigliucci states (Pigliucci 2012), "Drift has to do with stochastic events in generation-to-generation population sampling of gametes The strength of drift is inversely proportional to population size, which also means it has an antagonistic effect to natural selection (whose strength is directly proportional to population size). This leads to the drift-barrier for mutation rate evolution (Lynch *et al* 2016) and so to the views on the relation between drift and selection quoted in Section 1.

Drift is not a physical process in individual genes or organisms, but a statistical property of populations, and so is not strictly a causal effect (Millstein *et al* 2009, Matthen 2010, Lange 2013): it is analogous to kinematic rather than dynamical effects in physics. It is a result that necessarily follows from statistical relations, if one assumes the nature of the statistical process: usually that each copy of the gene in the new generation is drawn randomly and independently from the genes in the previous generation.

### 3.3 Multiple Realisability and Equivalence classes

A key issue is the vast degeneracy of the genotype ➔ phenotype map **BU1,** that is, the **multiple realisability** of phenotypes at level **L2** via genotypes at level **L1**.

This is a crucial property, labelled by Carroll (2008) as the functional equivalence of distant orthologs and paralogs. It is what makes a search of the genotype space possible in the time available since life began (Wagner 2011, 2017) because it underlies the fact the large genetic change can leave phenotypes unchanged while variation takes place, through whatever mechanism. This is what enables search of the `adjacent possible' (Kauffman 1992, 1993) in the available time, as described in detail by Wagner.

It then plays a key role in the top-down realisation **TD1** of selection taking place at level **L2** at level **L1** as follows: a hallmark that top-down effects are occurring is when higher level entities or functions are realised via **equivalence classes** of lower level entities or functions (Pereboom and Kornblith 1991, Rickles 2006, Auletta *et al* 2008, Ellis 2016). It is this multiple realisability at lower levels of higher level structure and function that is at the core of emergence of irreducible new properties at higher levels (Rickles 2006, Ellis 2016, Bickle 2016, Jaworski 2018), because it is only at that higher level that the core causal pattern emerges. It is not possible to specify the action at a lower level in a way that reveals the causal patterns that drive what is happening; in philosophical terms, the attempt to do so results in descriptions that are not natural kinds (Murphy and Brown 2005; Fodor, see Bedau and Humphrey 2008:403-407; Ellis 2016:374).

Descriptions that are natural kinds ( "The eagle swooped down in order to catch its prey") emerge at higher levels. This higher level action is enabled by lower level causation shaped so as to result in the desired higher level action (in this case, flows of electrons in the eagle's wing muscles). Existence of equivalence classes of processes at the lower level (all the billions of different ways electrons can flow in the eagle's muscles to achieve the same result) characterise when top-down causation enabling higher level function is taking place (Auletta *et al* 2012, Ellis 2016).

The same is true here of the top-down realisation **TD1**: the clear selection process at level **L2** is masked at level **L1** because it results in selection of equivalence classes of genotypes at that level, of which only one is realised in each selection event.

**3.4 Many more levels:**

There are many more levels of biological emergence than shown in Figure 2, which is a highly simplified representation of the real situation (Rhoades and Pflanzer 1989, Randall at al 2002, Campbell and Reece 2005. Noble 2012, Ellis 2016).

---

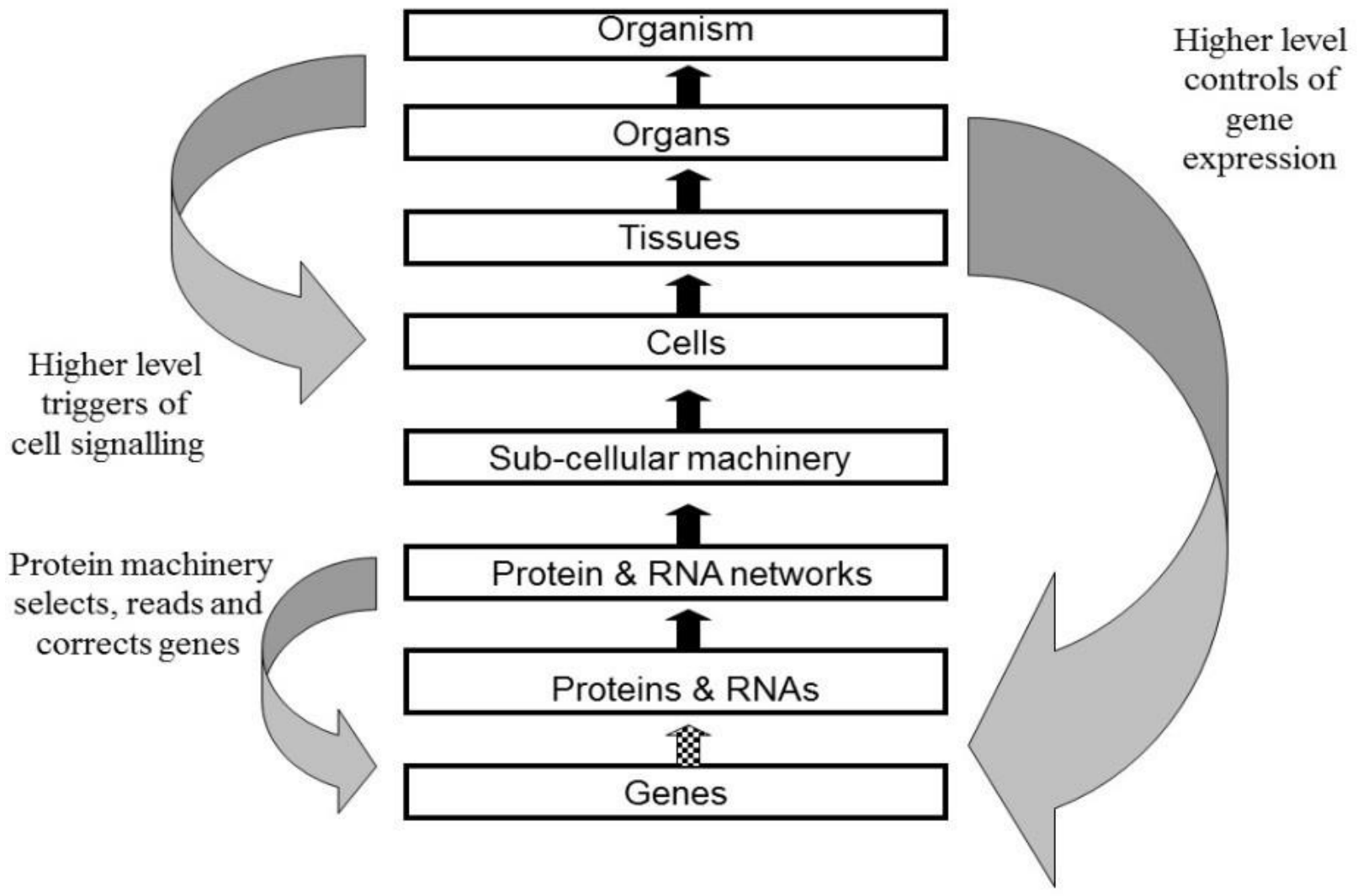

**Figure 3: Top-down physiological effects.** *Levels of emergence and associated top-down physiological control effects that are important functionally and during development. They will necessarily lead to a corresponding chain of top down effects during evolutionary selection. The cellular level is particularly important* From Noble (2012), with permission,

---

Figure 4 Indicates these levels. Both bottom up and top-down causation take place between each of these levels (Noble 2008, 2012, 2016, Ellis 2016), and what is represented in Figure 1 as a single top-down process **TD1** is in fact a chaining down of processes **TD1n** that takes place successively between many more levels **L1n** between level **L1** and **L2**. The top-down chaining of physiological effects is indicated in Figure 3; consequent downward chaining of selection effects will necessarily also occur during evolution, as these are the links from higher level necessary functions (such as the beating of the heart) to lower level processes that enable this to happen (such as gene regulation in the heart, see Noble 2002, 2012) to the genes that enable those lower level processes. In Figure 4 we give a simplified representation of this chaining through levels from **L2** to **L1** by including a level **L1A** of developmental systems (gene regulatory networks, metabolic networks, cell signalling networks) that is higher than the level **L1** of genes and proteins, because it is this level that determines developmental outcomes **BU1** through developmental processes as discussed in Section 4.1.

### 3.5 The negative view of natural selection and contrastive focus

The negative view suggests selection cannot explain the origin of individual traits (Pust 2001, Stegmann 2010, Martinez and Moya 2011, Birch 2012). Birch expresses the issue this way: "Can natural selection help explain why a particular organism has the traits it does? On the positive view of natural selection, it can: past selection for some trait can help explain why a later individual instantiates that trait. On the negative view of natural selection, it cannot: selection can explain the distribution and origin of trait types in a population, but it cannot explain the possession of any particular trait token by any particular individual .. [this]threatens to render inexplicable the adaptedness of an individual organism" Birch (2012). He responds by proposing a different criterion for explanatory relevance than implied by arguments for the negative view.

Pust (2001) nicely discusses it as an issue of **contrastive focus**, as in essence do Birch (2012) and Martinez and Moya (2011). The latter state that on the negative view, "natural selection is considered as a merely negative process, fulfilling only the function of filtering and distributing the percentage of variants already existing in a population, which originated and were shaped by other biological means". That is, the issue is if one focuses only on the selection step **S2** at level **L2** in Figure 2 (given by (2)), which is indeed a `negative process' although one that creates order out of disorder (Ellis 2016), or the entire multilevel process **NS1** given by.(1) leading to the full selection process **NS2** at level L2 given by (3), which is clearly a creative process (Darwin 1872, Mayr 2002, Carroll 2005).

To see the full constructive process one must take a multilevel view (Figure 2) which includes the randomisation process **RV1** leading to varied organisms **VO** ("Arrival of the Fittest", Wagner 2017) which can then be selected from to attain greater functionality. This multilevel process is a case of downward causation (Campbell 1974, Martinez and Moya 2011, Ellis 2017).

### 3.6 The overall result

The overall effect is that lower level components are selected by evolution so as to be adapted to higher level functions (Campbell 1974, Murphy and Brown 2007: 65-70, Ellis 2016, Ellis and Kopel 2017, Wagner 2017) as indicated in Figure2 and summarised in Eqns.(1)-(4). This provides an answer to the conundrum proposed in Section 1:

> **Proposed Resolution:** *Adaptation through selection is clear at the phenotype level **L2**, where it actually takes place, even if it is not clear at the genotype Level **L1**, where the effects that occur are a consequent result of those higher order processes and selection **TD1** takes place for equivalence classes of genes enabling higher level functions that are necessary for survival.*

This hierarchical structure is hidden if one takes a purely population genetics approach. The tensions noted by Birch (2016) are related to looking at the issue at these two different levels. For example, Natarajan et al (2016) show that the particular mutations in the different bird species living at high or low altitude seem to be random (or at least not predictable), but they all correlate with the same functional change in the phenotype.

## 4 Evo-Devo and the Extended Evolutionary Synthesis

The assumption of a 1-1 genotype to phenotype map is wrong in both directions, leading to a variety of complications encompassed in the **Evo-Devo viewpoint** (Carroll 2005, 2008). This incorporates specifically the importance of developmental systems (\S 4.1) and associated epigenetic effects (\S 4.2). This is developed further in the proposal of an **Extended Evolutionary Synthesis**, or ESS (Pigliucci and Müller 2010, c 2014, Laland *et al* 2015, Bateson *et al* 2017, Noble and Noble 2017), focusing on developmental bias, developmental plasticity, extra-genetic inheritance, and niche construction.

A key issue here is to what degree variational processes are random, given that developmental processes generate certain forms more readily than others (\S 4.3) Again many levels of emergence are involved, and this is discussed in the next section (Section 5), where *inter alia* the issue of niche construction is considered (\S 5.3).[6]

**4.1 Evo-Devo Effects**

The key point at issue here is that organisms are not generated by genes *per se*, but by developmental systems that determine what genes will get turned on when and where via epigenetic processes (Oyama *et al* 2001, Carroll 2008, Noble 2008, Gilbert and Epel 2009, Noble 2012, 2016). As traits are determined by such systems, when selection chains down from the organism level to the gene level (see Figure 2), it does so via the developmental systems level and specifically via **Gene Regulatory Networks** (GRNs), which therefore are a key target of selection. Consequently, we need as a minimum[7] to introduce a new level **L1A** of GRNs into Figure 2, leading to Figure 4. This is how developmental processes influence evolutionary processes, hence the label "Evo-Devo". One can think of the GRH level as a proxy for the crucial cellular level.

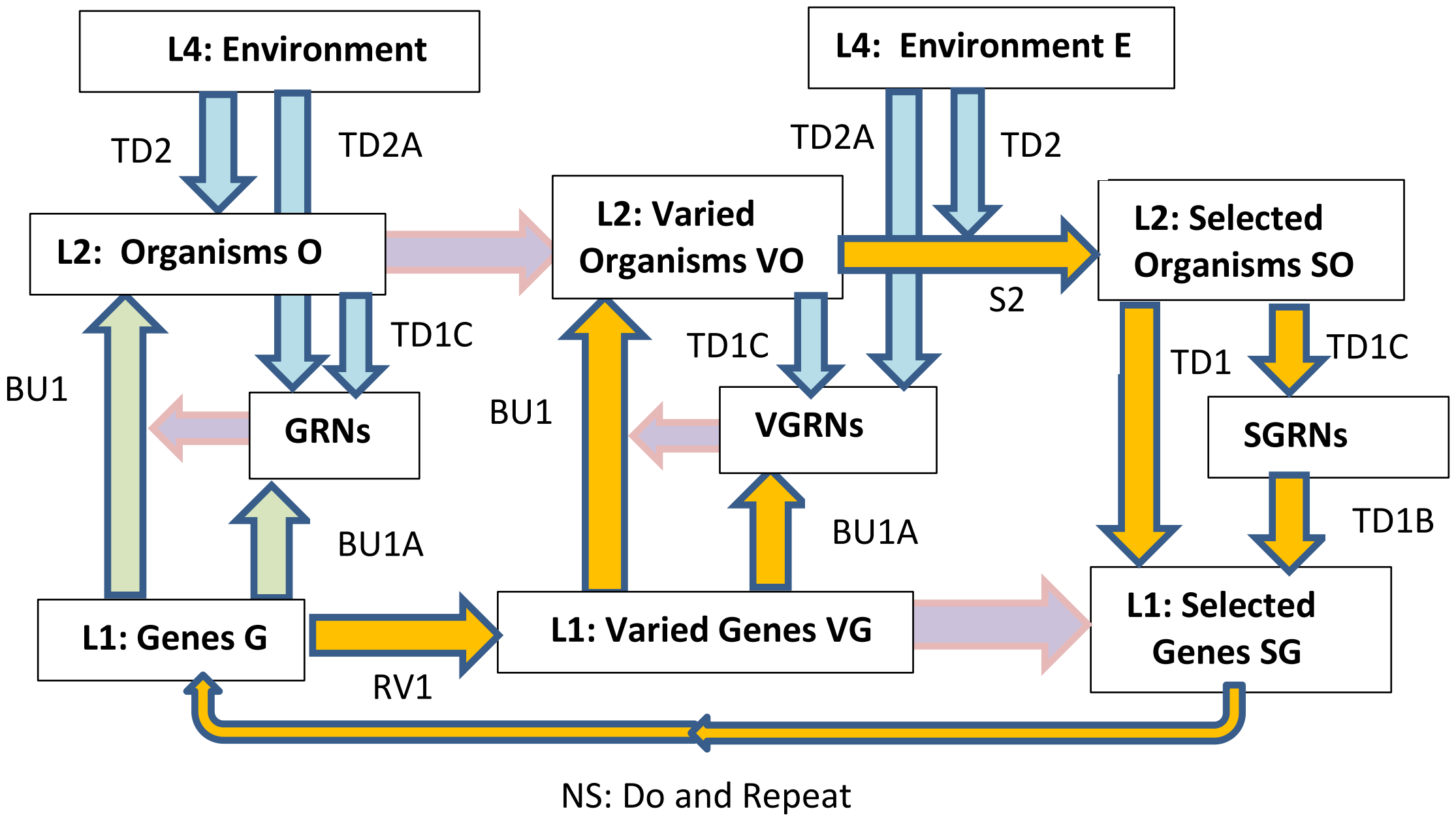

**Figure 4: The causal chain whereby selection leads to a new set of genes: gene regulatory networks.** *Extension of Figure 2 to take Evo-Devo effects into account.* *Not shown is the effect of individuals and the group on the environment (niche construction), see Figure 6.*

In more detail: Alberch (1991) introduced the genotype-phenotype "mapping function" **G→ P** which is much more complex than a one-to-one relation between genotype and phenotype (Gjuvsland *et al* 2013): the same phenotype may be obtained from different combinations of genetic informational resources (Pigliucci 2010) because of environmental plasticity (West Eberhard 1989, 2003), leading to Mosaic Pleiotropy (Carroll 2008), developmental bias, and developmental plasticity. The key question in developmental biology is what feature determines which genes are switched on where and when. This is controlled by **epigenetic processes** involving positional information (Wolpert 2002, Gilbert 2006), physiological information (Noble 2002, 2008, 2016), cell signalling, epistasis (the interaction between genes: Cordell 2002, Phillips 2008), and gene-environment interactions (Kauffman 1993, Cooper and Podlick 2002, Gilbert and Epel 2009).

[6] A further controversy arising in this broad context is the issue of whether there actually exists "junk DNA" or not (see e.g. comments by Larry Moran in his blog). I will not touch on that debate here.
[7] In principle, one should include all the levels shown in Figure 3 – but this would be very complex. Figure 4 is a workable compromise to illustrate the dynamical relations involved.

The term "epigenetics" however is contested and used in a variety of ways. Bird (2007) suggests the following unifying definition of **epigenetic events**: *the structural adaptation of chromosomal regions so as to register, signal or perpetuate altered activity states*. These are features not directly governed by the genetic code but by the cellular context, and include DNA methylation, covalent modification of histone proteins (Jaenisch and Bird 2003), and RNA mediated DNA regulation (Mattick 2001, 2012). They are enabled by Gene Regulatory Networks (Carroll 2008) which themselves are the product of developmental processes based on the genes coding for their proteins. These are key actors in what occurs (Carroll 2005, 2008, Mattick 2012, Noble 2008a, Noble *et al* 2014), and are also selected in order to give the desired developmental results, leading to genetic inheritance processes.

These multi-level evolutionary processes additionally select for metabolic networks, cell signalling networks, and proteins that enable the whole system to function (Wagner 2011, 2017). The basic structure is the same as Figure 2, but with new elements, see Figure 4.

- The GRNs control the bottom-up developmental processes **BU1** at each stage of the evolutionary cycle, that is first with the original genes **G** and then with the varied genes **VG.** This bottom-up process is indicated by the upward arrows in Figure 2.
- Positional information within the developing embryo and then in the organism (Wolpert 2002, Gilbert 2006) controls which genes get switched on where and when during development by the top-down epigenetic processes **TD1C,** and again occurs at each stage of the evolutionary cycle. Additionally, physiological information gets feed down to the GRNs all the time as part of the normal physiological functioning of the organism (Nobel 2008, 2016). This corresponds to the downward arrows in Figure 2 and to **TD1C** in Figure 4.
- Environmental factors influence this process by changing global parameters that then influence which genes get switched on and off (Carroll 2005, Gilbert and Epel 2009); this is the top-down process **TD2A**, again occurring at each stage of the developmental process.
- The top-down process **TD1C** leads to preferentially selected gene regulatory networks **SGRNs** which in turn lead to selection of preferred genotypes; that is the top-down process **TD1B**, which turns this whole process into an inter-level feedback loop over time (West-Eberhard 2003).
- The **principle of equivalence classes** (\S3.4) still holds: there are a great many ways that the higher level needs can be realised by lower level processes, GRNs, and genotypes.

Two interacting opposing elements enter here:

- **Developmental bias** (Maynard Smith *et al* 1985, Kauffman 1992, 1993) in the process **BU1** tends to channel development independent of the environment and leads to evolutionary convergence (Conway Morris 2003, McGhee 2011)
- **Developmental Plasticity** tends to channel development in response to environmental conditions via the process **TD2A,** channelling developmental outcomes as far as is allowed by developmental bias**.** Thus developmental outcomes respond to the environment, which is environmental plasticity (West Eberhard 1989, 2003)

**There is a robustness-evolvability tradeoff:** mutations affecting regulatory networks must both generate variance but also be tolerated at the phenotype level (Masel and Siegal 2009). Genotypes involving nonlinear dynamics allow expression levels to be robust to small perturbations, while generating high diversity (evolvability) under larger perturbations. Thus, nonlinearity breaks the robustness-evolvability trade-off in gene expression levels by allowing disparate responses to different mutations (Steinacher *et al*, 2016). This robustness promotes evolvability (Masel and Trotter 2010), which is crucially enables by the huge degeneracy of the genotype to phenotype map which allows exploration of genotype space in the time available since life began (Wagner 2011, 2017).

**4.2 Epigenetic Inheritance Processes:**

Organisms transmit more than genes across generations, namely as well as the genome, the next generation inherits all the basic cellular machinery that underlies cell function and sets the context for reading the genetic code (Alberts *et al* 2007, Hofmeyr 2017). Noble (2008a) states this as follows: "*There are two components to molecular inheritance: the genome DNA, which can be viewed as digital information, and the cellular machinery, which can, perhaps by contrast, be viewed as analogue information. … The egg cell machinery is just as molecular as the DNA. .. Both are used to enable the organism to capture and build the new molecules that enable it to develop, but the process involves a coding step in the case of DNA and proteins, while no such step is involved in the rest of the molecular inheritance. This is the essential difference*."

This may thus be referred to as **Extra-genetic inheritance**. It can include the transmission of epigenetic marks (Bird 2002, Jaenisch and Bird 2003, Laland ,Uller, *et al* 2014, Noble *et al* 2014) because features such as DNA methylation and covalent modification of histone proteins can sometimes also be inherited and alter DNA expression, and can thereby influence fertility, longevity and disease resistance (Laland, Uller, *et al* 2014, Heard and Martienssen 2014). For a specific example see Hu and Albertson (2017).

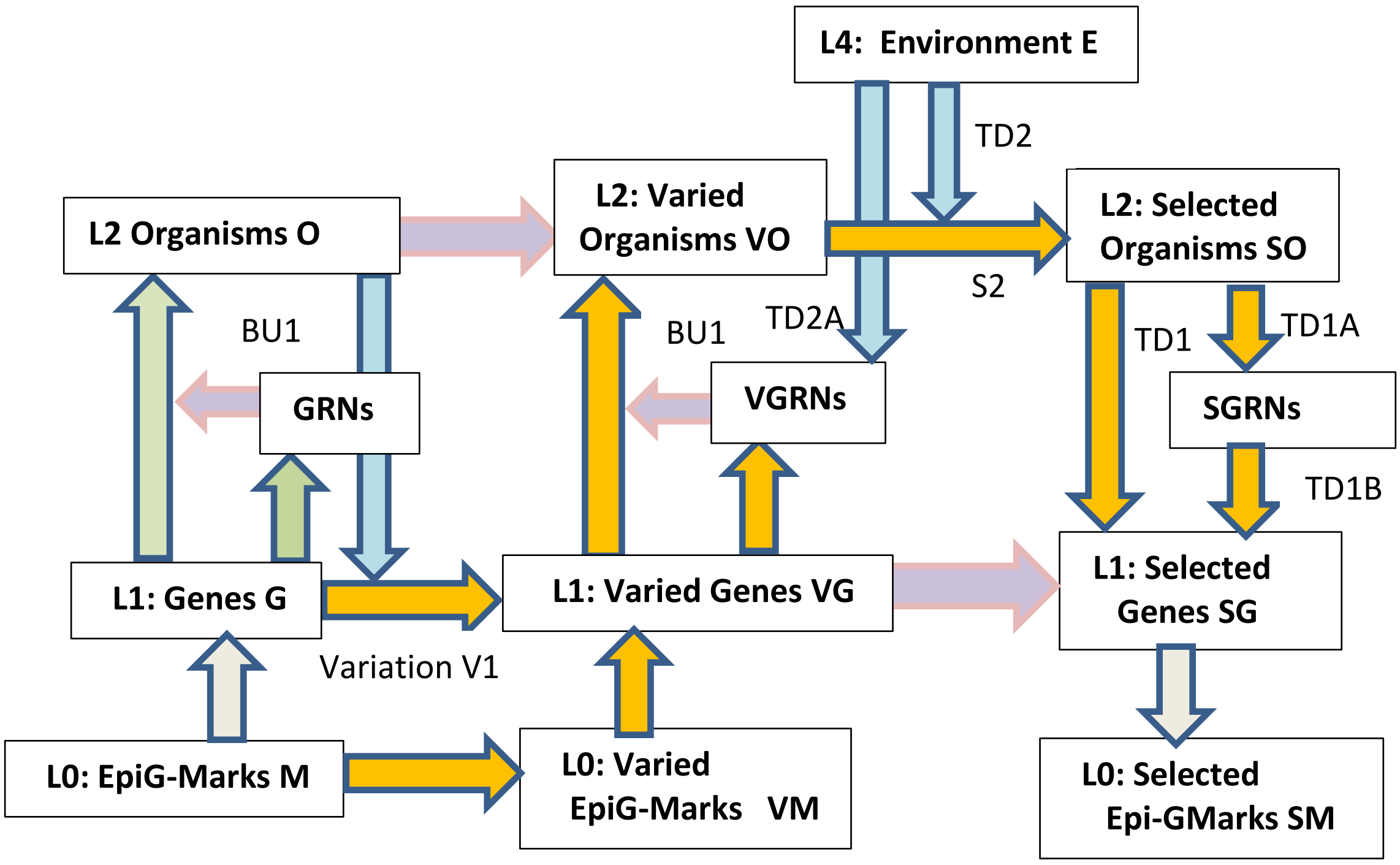

**Figure 5: The causal chain whereby selection leads to a new set of genetic contexts: extragenetic inheritance.** *Transmission of epigenetic effects to the next generation (here collectively labelled as "Epigenetic Marks"* **M***) will alter the way genes are read and so affect developmental outcomes.*

Figure 5 attempts to indicate this. Epigenetic Marks **M** at level **L0** lead to varied epigenetic marks **VM** in the next generation, which then affect reading of varied genes **VG** at level **L1** and hence developmental outcomes **VO** at level **L2**. The selection process **VO→ SO** will then chain down to lead to selection of both genes **SG** and epigenetic marks **SM** that tend to produce favourable outcomes at level **L1**. The selected epigenetic marks may or may not get passed on to the next generation (Dupont *et al* 2009); they may alternatively be "reset", and so not be passed on. The degree to which this is important is the subject of ongoing debate (Richards *et al* 2010, Laland, Uller, *et al* 2014, Wray *et al* 2014); Mattick (2012) suggests it is far more important than usually realised..

The phrase "epigenetic inheritance" is also used to cover cultural inheritance, allowing niche construction: that is not represented here, and is discussed in \S5.3.

### 4.3 Non-Random Variation

As mentioned already, much variation is not random at the level **L1** because developmental processes generate certain forms more readily than others (Kauffmann 1992, 1993, Laland Uller *et al* 2014). Such developmental bias can be regarded as a creative element leading to convergent evolution at that level (Conway Morris 2003, McGhee 2011, Laland, Uller, *et al* 2014).

However additionally, Noble and Noble (2017) point out that organisms and their interacting populations have evolved guided random mutational mechanisms (`hypermutation', Odegard and Schatz 2006) by which they can harness blind stochasticity and so generate rapid functional responses to environmental challenges by re-organising their genomes and/or their regulatory networks. Epigenetic as well as DNA changes are involved, the key feature being specific targeting of the relevant part of the genome due to higher-level buffering by regulatory networks in addition to differential genome mutation rates. As stated by Noble and Noble, "the guidance does not lie at the genome level…. At the genome level the process appears blind, it depends on stochastic mutation. The functionality enabling the process to be described as guided lies in the system as a whole".

### 4.4 The Overall Effect

The overall result of taking these additional effects into account is that the basic view put in the last section, summarised in Section 3.7 remains unchanged:

> **The overall effect** *is that adaptation through selection is clear at the phenotype level* ***L2****, where it actually takes place, even if it is not clear at the genotype Level* ***L1*** *(Figures 4 and 5) as summarised in Eqns. (1)-(4).*

Indeed that view is strengthened because the further mechanisms identified here lead to even larger equivalence classes at lower levels corresponding to higher level structures and functions.

## 5 Multilevel selection, kin selection, and niche construction

The above is of course a multi-level approach. But it has not touched on the vast literature around the issues of evolution of eusociality and altruism, and kin selection versus group selection: see Lewontin (1970), Maynard-Smith and Price (1973), Maynard Smith (1982), Alonso (1998), Okasha (2006), Wade *et al* (2010), Pigliucci (2010a) , Marshall (2011), Martinez and Moya (2011), Birch and Okasha (2015), Kramer and Meunier (2016),Birch (2018), and references therein.

**5.1 Multilevel Selection** The relevant emergent levels are displayed in Figure 6. From a macro-level causal viewpoint, there are two distinct things that can happen.

- First, there are clear cases where individual organisms **O** at level **L2** have higher survival value than other individual organisms in the ecosystem: they may be stronger or faster or able to fly or have better vision. Their preferential selection is the top-down effect **TD2** to the organism level. In those cases, a population **P** at level **L3** comprised of such organisms will be more likely to survive than populations comprised of organisms that are less fit individually. Thus the upward emergence process **BU2** then confers greater survival value on the population **P** at level **L3**. This effect (Walsh et al 2017) relies on linearity: the population is just the sum of its parts. That preferential survival value at the organism level **L2** will chain down to the gene level **L1** by the top-down process **TD1**, as before.

- Second, when genuine emergence of higher level traits occur, populations are not just aggregates of lower level trains (Ellis 2013)[8]. There are clear cases where the whole is more than just the sum of its parts, and an emergent trait at the population level **L3** has survival value for the population itself relative to other populations (Lehmann *et al*, 2007). This is the case for instance as regards development of social systems enabled by language, resulting in technology (weapons, transport, controlled energy use, and so on) that enhances survival and cannot be developed by individuals acting on their own. This can provide much greater survival value for the population **P** as a whole than if its constituent organisms just lived in the same area without the social interactions leading to these social innovations.

This results in the top-down process of group selection **TD4**, leading in turn to enhanced survival rates for organisms **O** belonging to that group. This is the top-down process **TD3** chaining selection down to level **L2** from the enhanced group survival rates at level **L3**. As before, this will chain on down via the process **TD1** to the level of developmental systems and genes.

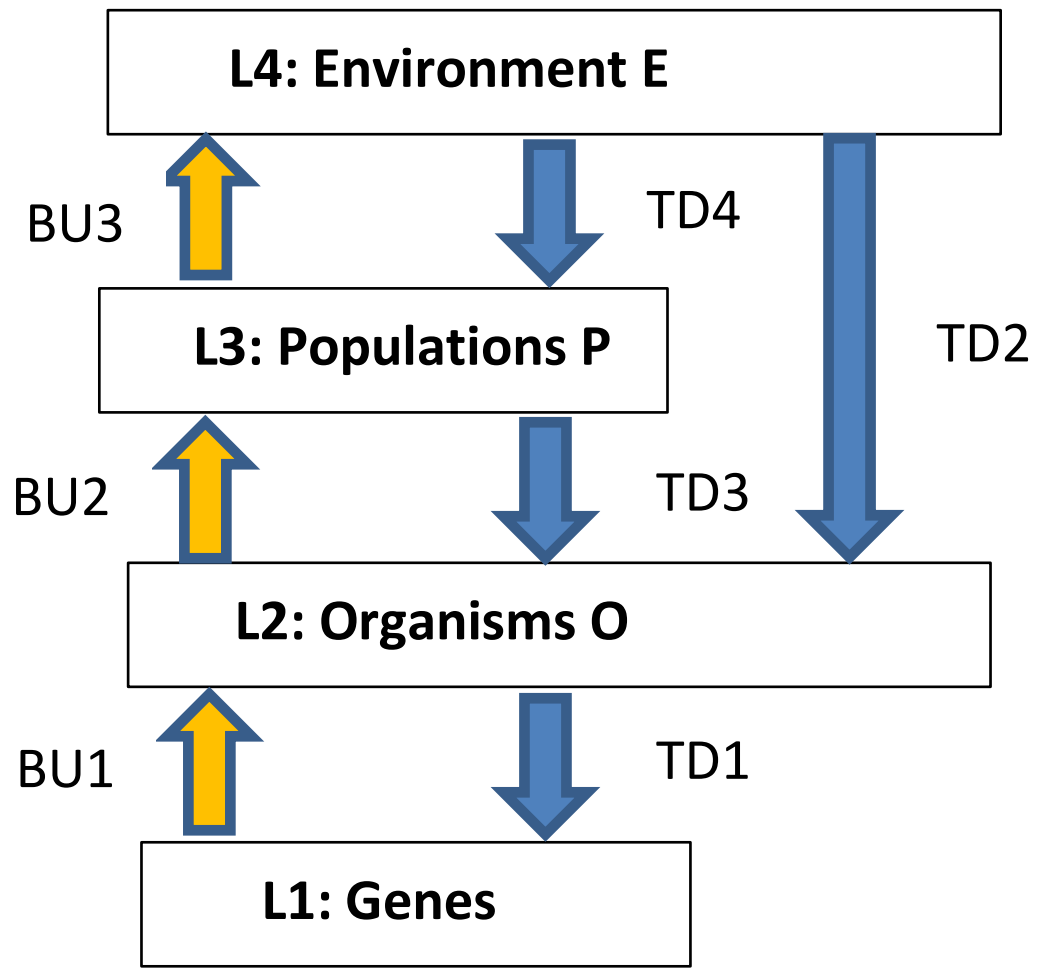

**Figure 6: The broad levels involved in evolutionary selection between groups and individuals.** *The effects of the environment included. Not shown is the direct effect of the individual on the environment, because the group effect is so much larger.*

This is kin selection in the following sense: referring to Traulsen and Nowak (2006), Lehmann *et al* (2007) say "*It is quite obvious that the mechanism that allows cooperation to evolve under T&N's life cycle is kin selection; interactions occur within groups, and individuals from the same group are related (i.e., they share a more recent common ancestor than individuals sampled randomly from the*

[8] For a graphic demonstration of the protective advantage provided by emergence of the common purpose of a collective of buffalo, see https://www.youtube.com/watch?v=LU8DDYz68kM. The whole is manifestly greater than the parts because it acts as a collective.

*whole population). Hence, T&N's model falls into the scope of Hamilton's inclusive fitness theory,*[9] *which is a general method for analyzing selection*". This form of kin selection occurs in the multilevel context represented by Figure 6 through **TD3** combined with **TD1**.

### 5.2 Kin selection/inclusive fitness versus group selection/multilevel selection

A key issue arises here: Why are altruistic individuals selected for, so promoting group survival but usually at their own expense, rather than selfish individuals being selected for**,** because they promote their own welfare at the expense of the group? How does one deal with the Free Loader problem? This issue is often characterised as a tension between *Kin selection,* the preferential selection (at Level **L2**) because of inclusive fitness of individuals who happen to be related, and *multilevel selection* or group selection, which is preferential selection (at **Level L3**) of the group *per se* when "competition occurs between groups *sensu stricto"* (Lehmann *et al* 2007)**.**
Birch and Okasha (2015) state this as follows:

> ***Kin selection** emphasizes the relatedness between social partners as the crucial factor mediating the spread of a prosocial behavior. **Multilevel selection**, in contrast, emphasizes the interplay of selection within groups and between groups … Within any group, altruists will be at a selective disadvantage vis á vis their selfish counterparts, but groups containing a high proportion of altruists may outcompete groups containing a lower proportion. So, for an altruistic behavior to spread, the between-group component of selection must trump the within-group component*

There is a large literature on the contest between within-group and intra-group selection, and which is most important in a particular context; for a summary, see Birch and Okasha (2015). Mathematically, this relates to Hamilton's Theory (Hamilton 1964) and the Price equation (Price 1970).

**Hamilton's Theory** The theory of kin selection (Hamilton's 1964) is the best-known framework for understanding the evolution of social behaviour. According to Birch and Okasha (2015),

> *"The basic empirical prediction of kin selection theory is that social behavior should correlate with genetic relatedness; in particular, altruistic actions, which are costly to the actor but benefit others, are more likely to be directed toward relatives. This qualitative prediction has been amply confirmed in diverse taxa … But despite its empirical success, kin selection theory is not without its critics".*

In fact major conflicts have raged around Hamilton's equation. Birch and Okasha (2015) claim that different protagonists in this conflict are using different versions of Hamilton's rule. To resolve it, they propose distinguishing three versions of this rule: Special (*HRS*), general (*HRG*), and approximate (*HRA*). *HRS* is an exact result for any model with an additive payoff structure. It therefore cannot represent emergent properties, where the whole is more than the sum of the parts and so won't represent group selection effects. *HRG* is an exact version of the rule that remains correct no matter how complicated the payoff structure may be, because all relevant payoff parameters are implicitly taken into account in the calculation of the costs and benefit; "*we are abstracting away from the complex causal details of social interaction to focus on the overarching statistical relationship between genotype and fitness.*" Debate about the usefulness of HRG is ongoing. *HRA* sacrifices a degree of this generality, relying on the idea that, when selection is weak and gene action is additive, a first-order approximation that neglects deviations from payoff additivity is justified When this is valid is clearly context dependent. This approximate version is the one most commonly used by kin selection theorists. But it does not obviously have a multilevel structure.

[9] The Inclusive Fitness controversy is dealt with in Birch and Okasha (2013) and Birch (2018).

**Kin selection = group selection**?? Despite much contestation to the contrary, kin and multilevel selection are often regarded as equivalent The mathematical formalisms are the same (Lehmann at al 2007, Marshall 2011, Wade *et al* 2010, Frank 2013, Birch and Okasha 2015, Kramer and Meunier 2016), the difference between the approaches is simply a question of collecting terms in the Price equation in different ways. As stated by Birch and Okasha (2015),

> *"In any group-structured population, the total evolutionary change can be decomposed using either the kin selection partition or the multilevel partition. Moreover, it is easy to see that the kin selection criterion for the spread of a prosocial trait (rb > c) will be satisfied if and only if the multilevel criterion is satisfied (i.e., the covariance between groups is greater than that within groups). Therefore, the two approaches are formally equivalent. .... Evolutionary biology, as are other sciences, is interested in constructing causal explanations; ideally, we want our descriptions of evolutionary change to capture the causal structure of the underlying selection process, as well as correctly computing allele frequency change. So, although kin and multilevel selection may be formally equivalent, it does not follow that they are also equally good as causal representations."*

The issue is which best represents the causal situation. That depends on context, and the context is the multilevel situation shown in Figure 6, together with the knowledge that group selection has indeed taken place in some contexts in the past. To look at this, I consider relevant causal mechanisms.

### 5.3 Through what mechanisms does multilevel selection occur?

There are basically two contrasting views on mechanisms underlying the social interactions that drive social evolution: the *Selfish Individual Hypothesis* and the *Social Brain Hypothesis*.

**The Selfish Individual hypothesis:** The assumption underlying much evolutionary analyses is the assumption of standard economics that individuals will always behave in a selfish way. For example Lewontin (1970) writes,

> "*In a regulated population it is clearly advantageous for a single individual to depart from cooperative behavior, just as it is advantageous for the single depositor to run to the bank in the economic analogy*".

This understanding is formalised in **Evolutionary game theory** (Maynard Smith and Price 1973, Maynard Smith 1982).[10] But that view is based on a rational expectations model of selfish behaviour – the standard economics assumption - that is not a realistic characterisation of how most people actually behave in society. It may possibly represent laboratory results where real-world complexities such as social obligations have been removed in order to get replicable results. But in fact families exist, social communities exist, and kindness and courtesy are common in the real world,

**The Social Brain hypothesis,** that the large brains of primates reflect the computational demands of their complex social systems (Dunbar 1998, 2003, 2014), is supported by much data. It assumes that primates and specifically humans mainly live communally in social groups instead of being scattered around as hermits – which indeed they do. This structuring is because at the population level, coherent social groups, with language and developed technology as well as community spirit, have a major competitive advantage over mere aggregations of isolated individuals who cannot call on group resources to solve problems[11]. Selection therefore takes place for individuals at level **L1** whose nature is such as to lead to existence of coherent social groups at level **L2:** the selective advantage of such groups at level **L2** has chained down to so as to preferentially select individuals at level **L1** with social brains (Dunbar 1998, 2003, 2014).

---

[10] There is a good summary in Wikipedia: https://en.wikipedia.org/wiki/Evolutionary_game_theory.
[11] For a graphic example, see the barn raising video at https://www.youtube.com/watch?v=y1CPO4R8o5M.

The undoubted existence of social groups in the animal kingdom is direct evidence for selection at the group level **L2** for such coherent social groups. Chaining down to the individual level **L1**, there is preferential selection for individuals at that level who tend to create emergent groups at level **L2**. This requires a social brain, preferring living in social groups where cooperation takes place.

**Is there a biological mechanism that leads to a social brain?** Sober and Sloan Wilson (1998) discuss the evolution and psychology of unselfish behaviour, proposing mechanisms of psychological altruism. The view I propose is that some of the innate primary affective systems (Panksepp 1998, Panksepp and Biven 2012, Ellis and Solms 2018) have been developed precisely in order to lead to group formation and stabilisation. Specifically (Stevens and Price 2000, Toronchuk and Ellis 2012):

- A primary emotional system that makes us strongly wish to belong to a group and feel allegiance to it, and so underlies the social brain and a tendency to share resources, and
- A ranking/dominance primary emotional system that regulates group conflict about resource sharing once a group has formed.

Nature has found it so important for survival that these innate mechanisms have been developed through evolutionary processes to strengthen the cooperation of individuals in societies. That is, they exist because of top-down selection from the group level **L3** to the level **L2** of individuals who have these traits, to the level **L1** of genetic and developmental systems that will ensure that they exist (Ellis and Solms 2017). Furthermore, they are reinforced by

- Social norms establishing what is acceptable in a society, reinforced by punishment (Boyd and Richerson 1992)
- Secondary emotions such as guilt and shame that strongly reinforce holding to the established social norms of a society.

Together these mechanisms largely negate the kind of logic represented by the Prisoner's Dilemma and similar analyses of social interactions. Dunbar's ideas are well supported by evidence, and should be taken as important pointers to what is really going on:

> **The social brain and social groups:** *humans and other higher animals do indeed have social brains, and have been selected to have them because they make possible major benefits arising from living in coherent social groups. This can only happen by multilevel selection from the group level down to the individual level and then down to the development systems level.*

### 5.4 Niche construction BU3

Niche construction effects **BU3** are all the processes whereby organisms alter the environment (Laland *et al* 2000, O'Brien and Laland 2012, Laland, Odling Smee and Turner 2014), so altering the selective context within which evolutionary processes take place. Thus this in turn affects the top-down processes **TD4** and **TD2**.

How important is this? It is crucial in some cases, for example. the Great Oxygenation event that transformed the Earth's atmosphere (Margulis and Sagan 1986). The development of agriculture is a key example of niche construction on the part of human beings (Laland *et al* 2000, Laland 2017). It also encompasses those structures and altered conditions that organisms leave to their descendants through their niche construction (Laland, Uller, *et al* 2014) — from beavers' dams to worm-processed soils to cities, technology (machine tools, computer systems, etc), and global climate change.

### 5.5 Many more levels, and Major Evolutionary Transitions

As mentioned in the previous sections, there are in fact many more levels of organisation than shown in Figure 6. This is indicated in Figure 3. Selective choice has chained down between all these levels, so that the lower level developmental systems will reliably lead to the simultaneous existence of each of the higher levels. This is necessary in order that they exist, as they demonstrably do (Rhoades and Pflanzer 1989, Randall *et al* 2002, Campbell and Reece 2005). A key feature then arises: how did these levels come into existence over the course of evolutionary history?

**Major evolutionary transitions:** Natural selection has led to the emergence of new levels of structure during evolutionary history (Maynard Smith and Szathmáry 1995, Calcott and Sterelny 2011, Szathmáry 2015), e.g.

- The emergence of Eukaryotes from Prokaryotes, after which mitochondria and chloroplasts can only replicate within a host cell;
- The emergence of multicellular organisms from single cell organisms, after which the relevant single cells are no longer viable as entities on their own;
- The origin of sex, after which organisms can only replicate as part of a sexual population;
- The origin of social groups, so that individual organisms such as ants, bees, wasps and termites can survive and pass on their genes only as part of a social group;
- The origin of human culture (Boyd and Richerson 1985, 2005), with symbolic systems such as language allowing emergence of social structures and technology that vastly increased competitive advantage *vis a vis* other groups.

This could not happen unless the evolutionary selective processes acting favoured emergence of the new higher levels, and then systematically acted downward from this higher level to the level of developmental processes that lead to its existence. Such emergence faces the same issue of within group and between group selection as discussed above (\S5.2):

> *"Why did not natural selection acting on entities at the lower level (replicating molecules, free-living prokaryotes, asexual protists, single cells, individual organisms) disrupt integration at the higher level (chromosomes, eukaryotic cells, sexual species, multicellular organisms, societies)*? (Maynard Smith and Szathmáry 1995:7)

Maynard Smith and Szathmáry (1995:8-10) see this as solved by

- **Immediate selective advantage** to individual replicators at the genetic level with a high degree of genetic relatedness between the units that combine in the higher organism. This is Hamilton's principle of kin selection.
- **Contingent irreversibility**: a ratchet effect whereby if an entity has replicated as part of a larger whole for a long time, it may have lost the capacity for independent replication it once had, for accidental reasons that have little to do with the evolution of the higher level entity in the first place
- **Central control**: the maintenance of organisations depends on some kind of central control. This is the principle of top-down causation discussed in Noble (2008,2012,2016) and Ellis (2012, 2016).

It leads to

- **The division of labour,** whereby an initially identical set of objects becomes differentiated and functionally specialised (the *principle of modularity*: Ellis 2016), which happened through developmental processes that have been selected for by evolution (Maynard Smith and Szathmáry 1995:12-13)
- **New ways of transmitting information**, including the origin of the genetic code, the origin of translation and encoded protein synthesis, the evolution of epigenetic inheritance with unlimited heredity leading to the emergences of plants, animals, and fungi, and the evolution of human language with unlimited semantic representation (Maynard Smith and Szathmáry 1995:13-14).

All of this involves top-down causation as indicated in Figure 6. As stated by Walker *et al*,

> "*Here we propose that a transition from bottom-up to top-down causation – mediated by a reversal in the flow of information from lower to higher levels of organization, to that from* *higher to lower levels of organization – is a driving force for most major evolutionary transitions. We suggest that many major evolutionary transitions might therefore be marked by a transition in causal structure"* (Walker *et al* 2012).

That is, whatever theories of competition within and between populations may say, it is a historical fact that higher levels of organisation have indeed emerged. This is only possible if top-down selective processes take place that lead to emergence of the new higher level. This cannot be captured by any equations that refer to a fixed number of levels. These crucial causal processes are hidden by any approach that implies the genotype level alone is the key to what is going on.

### 5.6 The overall effect

The outcome of the view put here is that

> **The overall result**: *One cannot adequately understand the processes in operation in evolutionary selection without taking a multi-level viewpoint (Figure 6). One can then discuss within that framework whether* **BU2**: *individual selection chaining up to the group level, or* **TD3**: *group selection chaining down to the individual level, are more important in specific cases. Both are important in different circumstances. In particular, selection at the group level is the key process leading to emergence of novelty during the evolutionary process*

Any equations used to describe these processes must be adequate to represent these multi-level processes and options. If they do not do so, they must be replaced by more adequate equations.

## 6: The process as a whole: relation to controversies

Martinez and Moya (2011) state

> *In this paper, using a multilevel approach, we defend the positive role of natural selection in the generation of organismal form. Despite the currently widespread opinion that natural selection only plays a negative role in the evolution of form, we argue, in contrast, that the Darwinian factor is a crucial (but not exclusive) factor in morphological organization. Analyzing some classic arguments, we propose incorporating the notion of 'downward causation' into the concept of 'natural selection.' In our opinion, this kind of causation is fundamental to the operation of selection as a creative evolutionary process*

This paper supports that view. Stochasticity at lower levels is harnessed by organisms to generate functionality: the order originates at higher levels, which constrain the components at lower levels (Noble 2017). A view focused primarily on genes will miss this causal structure.

### 6.1 The process as a whole

Schaffner (1998) illuminatingly discusses the relevant issues, phrasing them in terms of eleven theses, supported by many references, as follows:

(1) *The nature-nurture distinction is outmoded and needs to be replaced by a seamless unification approach in which genes and environment are "interacting and inseparable shapers of development."*
(2) *The relation between genes and organisms is "many-many" and the existence of significant "developmental noise" (chance events during development) precludes both gene-to-organism trait predictability (including behavioral traits) and organism trait-to-gene inferences. Thus the outcome is emergent.*
(3) *Genes do not "contain" the "information" that is a blueprint for traits, rather information discernible in maturing organisms develops the information is the product of an ontogeny.*

(4) *DNA sequences have no fixed meaning,but are informational only in context.*
(5) *Characterizing genes as causes of traits reflects outmoded preformationist thinking. Genes do not even make neural structures in any direct way, they produce proteins that affect cell differentiation to yield neurons that become specific types of neurons in specific places with particular connections with other neurons.*
(6) *Developmental causation is not just "bottom up," but is also "top down." Genes are not the principal actors that produce traits (including behavioral traits), but are part of a complex system, in which the cytoplasm can influence the genes, extracellular hormones can influence the nucleus, external sensory stimulation can influence the genes, and the hormones can be influenced by the external environment.*
(7) *The most accurate way to describe trait development is to use the "norm of reaction" approach which "is a list or graph of the correspondence between different possible environments and the phenotypes that would result", but this does not yield deterministic predictions. Even norms of reactions have to have a temporal developmental dimension added to them.*
(8) *The classical ethology approach of Lorenz that distinguished between "learned" and "innate" behavior has to be replaced by an "interactionist," "epigenetic," "ecological," or "life cycle" approach.*
(9) *Classical behavioral genetics is also committed to a false nature-nurture dichotomy that mistakenly believes it can distinguish between the contributions of heredity and environment to behavior.*
(10) *An analysis of variance is not the same thing as an analysis of causes. Because classical behavioral genetics is a population-based discipline with its main method being analysis of variance, it can say nothing about the causes of individual development. Classical behavioral genetics thus can only address the question "how much of the variance" is "attributable to heredity and how much to environment," but not "how" hereditary and environment actually produce their effects.*
(11) *The concept of 'heritability' found at the core of classical behavioral genetics is generally useless and misleading: the nonadditivity of genetic effects will not permit its applicability except in highly specialized artificial circumstances.*

This is essentially the Extended Evolutionary Synthesis (EES) position, and accords well with the view put in this paper, summarised in Figure 7. The overall process reflects Mayr's "proximate/ ultimate causation" distinction (Mayr 1961, Laland *et al* 2011, Scholl and Pigliucci 2014, Pigliucci 2016): physiological and developmental effects occur on the one hand, and evolutionary effects on the other, both being crucial to outcomes in biology (Noble 2013, 2017).

### 6.2 The controversies

Proposed views of the controversies listed in Section 1 are as follows:

1. **Drift/selection**: The conflict between field biologists and physiologists ("functional biologists" or "evolutionary ecologists") on the one hand, and those working in molecular evolution ("evolutionary biologists" or "population geneticists") on the other (Mayr 1961, Birch 2016), with the latter (e.g. Lynch 2007) vehemently claiming that drift is almost always more important.

The view here (\S 3.6) is that selection takes place at the or group phenotype level, and then that drives selection at genotype level – within an equivalence class. The genetic level is just one causal level in this integrated view (Noble 2012), The selection process may well not be easily apparent at the level of genes; that is because that is not the level where selection happens (Figures 2,4,6 and 7). Top-down processes (Campbell 1974, Murphy and Brrown 2005, Ellis 2016) are key to how this works, as emphasized by Noble (2017):

*"Stochasticity is harnessed by organisms to generate functionality. Randomness does not, therefore, necessarily imply lack of function or 'blind chance' at higher levels. In this respect, biology must resemble physics in generating order from disorder. This fact is contrary to Schrödinger's idea of biology generating phenotypic order from* molecular-*level order, which inspired the central dogma of molecular biology. The order originates at higher levels, which constrain the components at lower levels. We now know that this includes the genome, which is controlled by patterns of transcription factors and various epigenetic and reorganization mechanisms. These processes can occur in response to environmental stress, so that the genome becomes 'a highly sensitive organ of the cell' (McClintock). …Blind chance is necessary, but the origin of functional variation is not at the molecular level."*

That is the view of this paper.

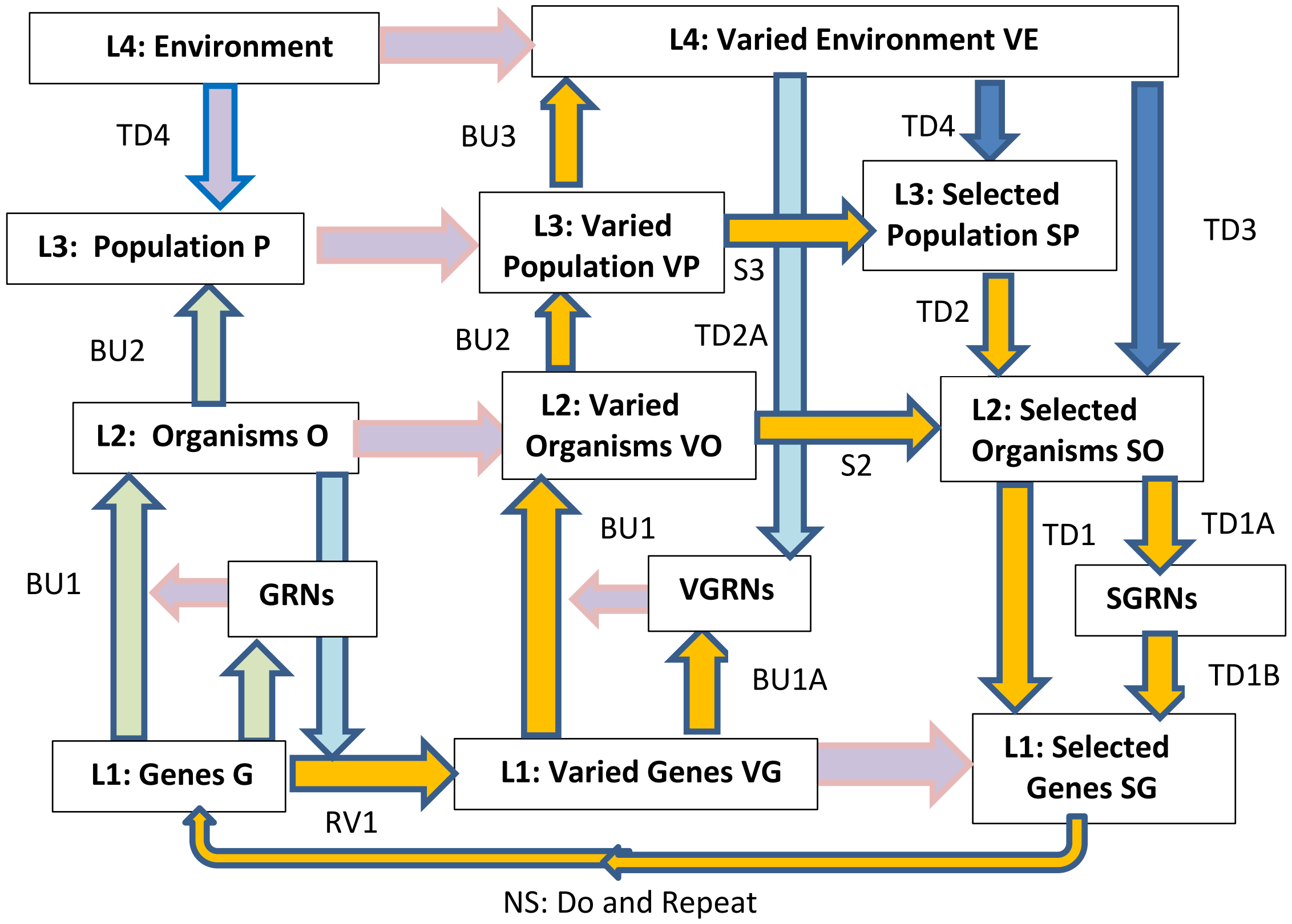

**Figure 7: The full causal chain whereby selection leads to a new set of genes (yellow arrows),** *influenced by the environment (blue arrows), but omitting the epigenetic marks.*

## 2. The negative view of selection

This is an issue of criteria for explanatory relevance (Pust 2001, Birch 2012), see \S3.5. Selection is not a negative process if one takes the whole dynamic process (Figures 2 and 7) into account.

## 3. The Extended Evolutionary Synthesis versus Standard Evolutionary Theory

The Extended Evolutionary Synthesis (EES: Pigliucci and Müller 2010, Laland *et al* 2015), which includes the Evo-Devo view (Carroll 2005, 2008, Gilbert and Epel 2009), emphasizes the interactions of developmental systems and physiology with evolutionary processes, and of evolutionary processes on the environment (Laland *et al* 2000, Laland *et al* 2014). This leads to developmental bias, developmental plasticity, extra-genetic inheritance (particularly culture, allowing niche construction), and epigenetic marks. The basic view of the previous section as regards the overall dynamics of selection remains unchanged (\S4.4)

What of the debate between Laland, Uller, *et al* (2014) proposing the EES should replace Standard Evolutionary Theory (SET), and Wray *et al* (2014) proposing the SET is just fine? The key point here in the end is the SET is ***a gene-centric view of evolution*** (Wray *et al* 2014) whereas the EES is ***one where genes are just one actor among many*** (Laland, Uller, *et al* 2014, Noble 2012, Noble 2017). Wray *et al* (2014) make the following core statement:

> *Diluting what Laland and colleagues deride as a 'gene-centric' view would de-emphasize the most powerfully predictive, broadly applicable and empirically validated component of evolutionary theory. Changes in the hereditary material are an essential part of adaptation and speciation. The precise genetic basis for countless adaptations has been documented in detail, ranging from antibiotic resistance in bacteria to camouflage coloration in deer mice, to lactose tolerance in humans.*

In contrast, Laland, Uller, *et al* (2014) state

> *In our view, this 'gene-centric' focus fails to capture the full gamut of processes that direct evolution. Missing pieces include how physical development influences the generation of variation (developmental bias); how the environment directly shapes organisms' traits (plasticity); how organisms modify environments (niche construction); and how organisms transmit more than genes across generations (extra-genetic inheritance). For SET, these phenomena are just outcomes of evolution. For the EES, they are also causes.*

The EES view is supported by the view put here. Important as they are, genes are just one of the players in what is happening; and they are not at the level where selective outcomes are decided (Figures 2 to 7). They are enablers rather than drivers of evolution.

## 4. The multilevel/kin selection debate

This debate is tied in to the levels of selection issue (Lewontin 1970, Okasha 2006). Kin selection does indeed take place, but it takes place in a multilevel context, and it causal structure cannot be understood outside that context, see \S5.6 and Wade *et al* (2010), Kramer and Meunier (2016). Selection often takes place at the group level, and in particular this is crucial for many major evolutionary transitions (Calcott and Sterelny 2011:9), which could not occur otherwise. The situation is stated clearly by Sloan Wilson and Sober (1994), who state

> *"We show that the rejection of group selection was based on a misplaced emphasis on genes as "replicators" which is in fact irrelevant to the question of whether groups can be like individuals in their functional organization. The fundamental question is whether social groups and other higher-level entities can be "vehicles" of selection. When this elementary fact is recognized, group selection emerges as an important force in nature and what seem to be competing theories, such as kin selection and reciprocity, reappear as special cases of group selection. The result is a unified theory of natural selection that operates on a nested hierarchy of units"*

Cohesive groups that can be objects of selection

- Will not exist *inter alia* in the cases of bacteria, fungi (including yeast), worms (including *c elegans*), flies (including *drosophila*), bird flocks, and fish schools. In these cases, the whole is just the sum of its parts.

- May exist in the cases of wales, buffalo, lions, wild dogs, and similar species that hunt communally, as well as elephants and some of the great apes. The cases of birds and bats are unclear.
- Clearly do exist in the case of ant colonies, eusocial insects (including bees) and animals (including naked mole rats), and social animals where a communal culture enhances survival prospects. The issue has to be decided on a case by case basis. But human beings are certainly amongst them.

### 6.5 Overall

The way adaptation is recognisable at higher levels is emphasize by Noble and Noble (2017), who state

> "*The targeted mechanism in the immune system has been known and intensely studied for many years . So, how did many people not realise that it is a physiologically guided process? The answer is that the guidance does not lie at the genome level. At the genome level the process appears blind. It depends on stochastic mutation. The functionality enabling the process to be described as guided lies in the system as a whole*"

A multilevel-view, allowing for the effects of top-down causation as indicated in Figures 2, 4, and 7, can reconcile the views of the population geneticists and field biologists. The former see adaptive effects swamped by random genetic drift, the latter see clear evidence of organisms being adapted to their environment. The point is that selection does not take place at the genome level, and there is a vast equivalence class of genotypes that correspond to the same genotype: that is why the clear adaptive processes apparent at levels **L2** and **L3** can get hidden at level **L1**.

Note that I have not claimed that rigorous evolutionary extremization principles exist for the situations considered here (cf. Metz et al 2008). Relative advantage as characterised by reproductive potential is what counts and leads to adaptive outcomes, as considered in studies such as those by Thygesen et al (2005) of optimal life histories for fishes or other animals in relation to the size spectrum of the ecological community in which they are both predators and prey. These studies are studies of the selective processes taking place at level **L2** and their dependence on properties at level **L3**. They assume, as this paper does, that adaptation does indeed take place, and studies the details of how it takes place and what quantities get optimized.

**Acknowledgements:**

I thank Denis Noble and Keith Farnsworth for helpful comments.